\documentclass[pra,twocolumn,amsmath,amssymb,superscriptaddress]{revtex4-2}

\usepackage[utf8]{inputenc}
\usepackage[english]{babel}

\usepackage{framed}
\usepackage{xcolor}

\usepackage{latexsym}
\usepackage{graphicx}
\usepackage{amsfonts,amssymb,amsmath}
\usepackage{hyperref}

\begin{document}

\title{Quantum Radars and Lidars: { Concepts, Realizations, Perspectives}}

\author{Gregory Ya. Slepyan}

\affiliation{School of Electrical Engineering,
Tel Aviv University,
Tel Aviv 69978, Israel (e-mail: slepyan@tauex.tau.ac.il and boag@tau.ac.il)}

\author{Svetlana Vlasenko}
\affiliation{B. I. Stepanov Institute of Physics, NAS of Belarus, Nezavisimosti ave. 68, 220072 Minsk, Belarus}

\author{Dmitri Mogilevtsev}

\affiliation{B. I. Stepanov Institute of Physics, NAS of Belarus, Nezavisimosti ave. 68, 220072 Minsk, Belarus}

\author{Amir Boag}

\affiliation{School of Electrical Engineering,
Tel Aviv University,
Tel Aviv 69978, Israel (e-mail: slepyan@tauex.tau.ac.il and boag@tau.ac.il)}

\begin{abstract}  Quantum radars and lidars are a novel, much-discussed,  and rapidly evolving ﬁeld of quantum science and technology, promising remarkable advantages in such basic tasks as target detection, ranging, and recognition. Quantum radars and lidars have already moved from the realm of theoretical considerations toward experiments, {{green}and, as is the case for lidars}, toward practical applications. Here, we review the underlying concepts and present basic configurations of quantum radars and lidars based on photonic entanglement and single-photon detection. We also brieﬂy discuss methods of producing entangled photons, such as spontaneous parametric down conversion, spontaneous four-wave mixing, Josephson parametric amplifier, and quantum antennas. { We show that quantum technologies open promising avenues toward} enhancement of signal-to-noise ratio and overcoming the Rayleigh limit in radar and lidar systems. 
\end{abstract}

\maketitle

\section{Introduction}

The so-called first quantum revolution started in the early 20th century and allowed scientists to understand the atomic structure and quantum nature of the electromagnetic field. The first quantum revolution provided the foundations for the development of many important devices such as transistors and lasers. The second quantum revolution started in the early 21st century. It opened the ways for manipulating and exploiting for practical purposes specifically quantum features, such as non-classical correlations between individual particles and non-classicality of states of the electromagnetic field. One of the most popular of these is the quantum entanglement, the “spooky action at a distance", which Einstein \textit{et al.} found disturbing as the fundamental principle of quantum theory \cite{einstein1935}. The second quantum revolution already brought significant development of quantum technologies and major technical advances in many diverse areas, such as quantum computing and communications \cite{nielsen_chuang_2010}, metrology, sensors, and imaging \cite{dowling2015}.

Quantum technologies jointly with nanotechnologies lead to a new field of engineering aimed at translating the effects of quantum physics into practical applications. Using solid-state quantum systems, atoms, ions, molecules, and photonic circuitry, one can design microscopic, but intrinsically quantum devices for generating, processing, emitting, and receiving quantum states \cite{Haroche2020FromCT,schuster,blais,carusotto,ourreview}. 

In this article, we review the basic concepts of quantum radars while highlighting the differences compared to their classical counterparts. We discuss how several concepts from the traditional radar technology \cite{Rich} (target detection sensitivity, noise resilience, and ranging accuracy) can be translated to quantum radars. We outline new far-field sensing protocols, examine engineering opportunities, enabled by their different realizations, and { highlight recent controversy arising around these realizations}. We consider the possibilities of achieving super-sensitivity and super-resolution using quantum correlations, and discuss the possibilities to create quantum correlated states in practice. In particular, we review the most commonly considered and relatively easy-to-create sources of entangled photon pairs both at microwave (Josephson parametric amplifier (JPA)) and optical frequencies (spontaneous parametric down conversion (SPDC), spontaneous four-wave mixing). The radar and its optical analogue (lidar) concepts are increasingly appreciated and used in the design of quantum-based far-field sensing devices. {Thus, we believe that the discussed theme holds a promise to become a prominent sub-field of quantum technologies in the coming years.}

\section{Quantum radars}
Radar is a system that uses electromagnetic waves (pulsed or continuous) to obtain information about an object (“target”) by detecting the field scattered by this object. The simplest radar task is the target detection or, in another words, the decision whether the target is present or absent inside the region of interest. Another type of the problem is to determine the target position and estimate its velocity. The most complicated task involves the recovery of the target configuration (imaging) via the analysis of its scattering pattern \cite{Rich}. 

Classical radars are able to successfully accomplish all these tasks. However, there are limitations dictated by sensitivity, noise, and losses of the detection systems, as well as noise present in the environment and leading to the existence of a threshold for signal-to-noise ratio (SNR), and the necessity to overcome it for making a functional radar. Also, there are fundamental limits dictated by the wave nature of the electromagnetic radiation which can be hardly overcome by increasing signal-to-noise ratio (especially, if one wants “stealth action”, i.e., low probabilities of signal interception).  For example, for the task of resolving two close independently emitting (or scattering) small objects by detecting intensity of the emitted field, the resolution is limited by the well-known relations between the wavelength of the field and the numerical aperture of the imaging set-up \cite{1880MNRAS..40..254R,Abbe2009}. So, when the details of the imaged target are smaller than the limiting value, one cannot resolve these details. The microscope-oriented version of this resolution criterion was suggested by Abbe in 1873 \cite{Abbe2009}; while the angular, far-field imaging oriented criterion was formulated by Lord Rayleigh in 1879 \cite{1880MNRAS..40..254R}. These criteria are based on rather empirical considerations, derived from the picture of overlapping diffraction patterns of the sources, and are not strict. Nevertheless, going far beyond the limits, imposed by these criteria, usually requires unfeasibly large expenditure of resources (measurement time, field intensity, etc.) \cite{Helstrom:70}. 

{ Notice that the resolution is much dependent on the statistical properties of the imaging field. For example, for the task of resolving two close independently emitting point sources, incoherent sources can give better resolution than coherent ones \cite{tsang2015}. Recently,  statistical information approach to the classical imaging  led to development of a new class of superior imaging methods \cite{tsang2019resolving}. In the middle  of the 20th century, use of correlations between emitted and returned signal were suggested for the so-called “noise radar” \cite{784050,Lukin_2001,Thayaparan2006NoiseRT}.  The noise radar produces a random signal, split into the transmitted beam and reference signal. The received field interferes with the delayed reference signal to produce the intensity correlation function. For the delay corresponding to the distance to the target, there is a sharp increase in the observed correlations. Because of the signal randomness, the noise radars were touted as having high immunity to noise and  low probability of intercept \cite{784050,Thayaparan2006NoiseRT}.

 Quantum correlations promise great advantages for imaging and sensing.} For example, quantum correlations even allow one to reach the ultimate quantum limit for measurement precision, the so called "Heisenberg limit" \cite{giovannetti2011advances}. The starting point for a widely growing interest in radar enhancement by the quantum correlations was created by a groundbreaking { and inspiring result} of Seth Lloyd in 2008. He invented the “Quantum Illumination” (QI) scheme \cite{Lloyd1463}.  This scheme promised huge enhancement of the target detection sensitivity despite the presence of strong noise background. Especially attractive was the possibility of obtaining this enhancement for very weak signal (at the level of a few photons), thus providing the possibility of stealth action. 

The last decade has seen intense development of quantum-enhanced far-field sensing and establishing connections with the other branches of quantum technologies, such as quantum metrology and imaging \cite{dowling,lanza,simon}. Initial theoretical developments (nicely summarized in the monograph \cite{lanza}) were followed by experiments in the microwave and optical frequency ranges \cite{zhang2015entanglement,lop,eng,luong2020,liu2019enhancing,zhang2020multidimensional,barz2015}, { and the advantage over the optimal classical scheme was demonstrated \cite{zhang2015entanglement}.   Recently, these experiments became a subject of intense controversy leading to the conclusion that for the moment there are no microwave QI schemes demonstrating an advantage over the optimal classical radars \cite{shapiro2020quantum,Cho1556,Jonsson2020ACB,p1,p2}.} 
{ The very possibility} of a realistic QI radar was doubted, and even in the case of such a radar being built it is  predicted to be impractically expensive \cite{9114614}. From the other side, QI protocol is certainly not the one and only possible way to enhance radars and lidars by quantumness. For example, there are such schemes as thresholded { optical} detection for enhancing sensitivity  \cite{cohen2019thresholded} and  quantum ranging \cite{maccone2020quantum}.  Also, such specifically quantum technological advancements as single-photon detectors have already established a firm place among lidar applications \cite{kirmani2014first, shin2016photon}. Even in their simplest ``click''/``no click'' version, they  enable a full reconstruction of a quantum state \cite{PhysRevLett.96.230401}. Single-photon detectors are being developed now for microwave fields \cite{kono,micro}.  It would be useful to mention that specific quantum  states designed to have low noise (so called "squeezed states") are  successfully used to strongly enhance sensitivity of gravitational waves detection \cite{PhysRevLett.123.231107}.

  In summary, despite the controversies, {  set-backs} and technological obstacles, for the moment the quantum-enhanced far-field sensing seems to be en route to developing practical highly sensitive { devices}, operating with low-intensity signals, providing low probabilities of intercept, and possibly achieving super-resolving target imaging.   

\section{Quantum concepts}

To explain the nature of possible quantum enhancement in far-field sensing, let us first briefly outline some fundamental principles of quantum theory. For discussing quantum radar implementations, we stick to a simple operational definition of “quantumness” of states as ability to provide observation results that cannot be reproduced with classical states for a specific measurement setup { (for example, having zero probability of coincident clicks on the detectors in Hong-Ou-Mandel interferometer with single-photon inputs \cite{PhysRevLett.59.2044}, or violating Bell inequalities \cite{PhysRevLett.23.880} ). Here, under the “classical states” we mean the ones obtained by arbitrary mixing of states that can be produced by maser or laser devices, or by classical current sources (i.e., positive-weight mixtures of coherent states).} For an introduction into fundamentals of quantum optics, please, see the paper by G. Hanson \cite{9096523}; for a more detailed discussion of electromagnetic field non-classicality, see, for example, a recent review \cite{doi:10.1116/1.5126696}.

In quantum mechanics, an observable physical quantity is described by the corresponding Hermitian operator. The state of the physical system is described by the wave-function, or, more generally, by the density matrix, which describes a statistical mixture of states corresponding to different wave-function. 
 Knowing the density matrix, one can predict the average results of all possible measurements performed with this state. An expected averaged measured value of any observable is given as the trace of the corresponding operator multiplied by the density matrix
 
 The nature of the quantum measurement is intrinsically stochastic. Repeating the measurement with the copies of the same quantum state will generally produce different results.
 So, any quantum measurement process gives an experimenter a sample of some probability distribution defined by the measurement setup and the state \cite{nielsen_chuang_2010}. 
 For example, with a simple ``bucket'' detector able to distinguish only between the presence or absence of signal, one can sample a binomial distribution (which is actively exploited for producing very high-quality random number generators, and some of them are already commercially available \cite{herrero2017quantum}).

Importantly, the quantumness can be manifested in lowering measurement noise below the classically attainable level.
For example, the field in the state with a particular integer number of photons (i.e., Fock state) will have maximally low intensity noise, and, as such, is a perfect candidate for estimation of transmission, which can be done with Fock states with much better precision than with coherent states \cite{braun2018quantum}.  { Another example is the already mentioned  squeezed states.  They are able to exhibit suppressed amplitude noise and  successfully used for reducing measurement noise  \cite{PhysRevLett.123.231107,barsotti2018squeezed}.  }

The modes of multi-mode field states can exhibit such specifically quantum correlations as entanglement, which is actively exploited in the quantum radar schemes for lowering noise. If the state of the multi-mode system is pure (i.e., can be described by just one wave-function), the entanglement manifests itself by inseparability. This implies that it is impossible to describe the evolution of every subsystem by its own independent wave-function, and the dynamics of the whole system as a direct product of individual wavefunctions.  For example, such situation arises when an emitter has equal possibility to create just $N$ photons into the spatially separated modes \textit{a} and \textit{b} \cite{dowling}.  The total state produced in that manner is named N00N state and is described by the wave function $|\Psi\rangle\propto|N\rangle_{\mathrm{a}}|0\rangle_{\mathrm{b}}+|0\rangle_{\mathrm{a}}|N\rangle_{\mathrm{b}}$, where $|N\rangle_{{\mathrm{a}},{\mathrm{b}}}$ are the $N$-photon Fock states and $|0\rangle_{{\mathrm{a}},{\mathrm{b}}}$ are the vacuum states of the modes \textit{a} and \textit{b}, respectively.  The entangled states are non-trivial to create and measure. One needs non-linear processes or emitters that produce no more than several photons in every emission act \cite{o2009photonic}. 

For a mixed state described by the density matrix, the determination of the entanglement presence is more complicated than for pure states \cite{horodecki2009quantum}. A mixed state can exhibit other kinds of correlations, which are weaker than the entanglement, but, nevertheless, are quantum. This is  referred to as a quantum discord, which is defined as a measure of non-classicality of information content of the state \cite{korolkova2019quantum}. In contrast to the notoriously fragile quantum entanglement (which can even disappear completely, or suffer a “sudden death” for a finite time under the actions of losses and added noise \cite{yu2009sudden}), the presence of discord can be very robust, and even be increased due to losses \cite{korolkova2019quantum}. The very ability of the quantum radar to outperform classical radars might be based on the resilience of the discord remaining after the initial entanglement is lost \cite{weedbrook2016discord}.

Often, one needs to build a quite involved  quantum measurement to realize the advantage offered by a quantum state (such as projection on the entangled state) and assess all the available information (as is actually the case for the quantum illumination \cite{dowling}). For example, a thresholded detection by filtering out low photon numbers,  has been recently suggested for quantum-enhanced lidar \cite{cohen2019thresholded}. 
 There is a way to check whether this or that kind of quantum measurement is sufficient to infer the information about the target. If one has measurements results, described by a complete set of $J$ probabilities, $p_j(x_1,\ldots,x_M )$, dependent on parameters $x_m$, which we want to infer, for the unbiased estimation, the variances of the estimated parameters are bounded from below by the Cram\'{e}r-Rao inequality 
\begin{equation}
\label{eqn:fisher}
    \operatorname{var}x_m\ge \frac{\left[F^{-1}\right]_{mm}}{N},\quad \left[F\right]_{mn}=\sum_j^J\frac{1}{p_j}\frac{\partial p_j}{\partial x_m}\frac{\partial p_j}{\partial x_n},
\end{equation}
which is well known and much used in statistical estimation theory \cite{kay1993fundamentals}. In expressions (\ref{eqn:fisher}), $F$ is the Fisher information matrix and $N$ is the number of the measurement runs or state copies (generally proportional to the measurement time). Quantum mechanics allows one to determine the quantum bound corresponding to the maximization of the Fisher information over all possible measurements. To do that, one needs to know only the quantum state \cite{petz2011introduction}. 
Inequality (\ref{eqn:fisher}) demonstrates another way to manifest quantumness through the measurement: the possibility to transcend a typical estimation error dependence on the number of runs, when this error changes as $1/\sqrt{N}$. This kind of dependence is known as “the standard quantum limit” (SQL). With entangled states such as the N00N state, one can reach the so-called “Heisenberg limit”, with the error changing as ${1}/{N}$, which is the maximal estimation precision allowed by quantum mechanics \cite{giovannetti2011advances}.

{ Notice that quantumness can be potentially manifested also by the difference in the way  the target back-scatters the field for quantum and classical illuminating states (i.e., between corresponding radar cross sections) \cite{lanza}. For example, this might take place when multi-photon scattering processes are significant (however, it is hardly the case for practical radar applications) \cite{lanzanew}.}  

\section{Quantum radar configurations}

Here, we consider quantum enhancement of radars for all three basic radar tasks: target detection, target ranging, and target imaging. For illustrating the concepts and their realizations, we discuss three schemes: { the already-famous much debated QI scheme} \cite{Lloyd1463}, and two more recent ones \cite{mikhalychev2018synthesis, maccone2020quantum}. All three schemes are based on the sources emitting entangled photons (Fig. \ref{fig:1}). 

\subsection{Quantum illumination for target detection}
QI scheme considers detection of a low  radar cross section (RCS) object in a thermal noise environment \cite{Lloyd1463}. {Let us demonstrate how the QI functions and why it did generate such interest as a means for quantum enhancement of far-field sensing. To that end, we consider a simple reference case of just a single photon illuminating a target, being reflected, and registered.}

\begin{figure}[h]
    \centering
    \includegraphics[width=0.45\textwidth]{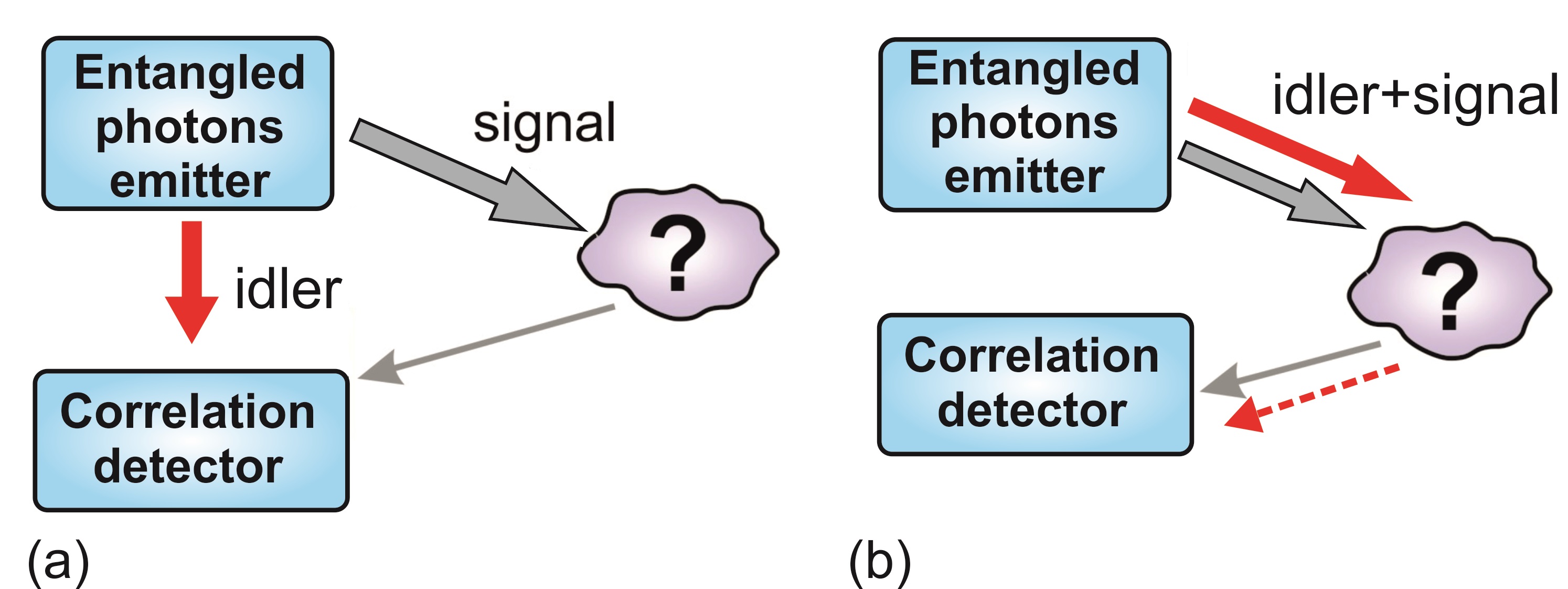}
    \caption{(a) Quantum illumination scheme for the target detection. Only a single beam of the entangled pair illuminated the target. (b) Quantum radar scheme for the ranging and resolving the target. All the entangled field illuminates the target. Correlation measurements on the returned field are performed.} 
    \label{fig:1}
\end{figure}

Let us assume that we have $d$ field modes, e.g. frequencies; our photon is somehow distributed among these modes and is described by the state $|\theta\rangle$. This photon is sent toward a weakly reflecting target. So, only a small portion,  $R\ll1$, of the sent photons returns to the receiving detector. On the way, this portion is mixed with a $d$-mode thermal state with $n$ photons per mode, while the total number of photons is small, $dn\ll1$. The receiver checks whether the returning photon is in the state $|\theta\rangle$ (projects the returning state on the state $|\theta\rangle$). Then, the probability to get a positive result for the target detection when the target is not actually present is  $p_{\mathrm{false}}\approx n$, and the probability to get the positive result in the presence of the target is $p_{\mathrm{true}}\approx n+R$ \cite{Lloyd1463}. These results are quite intuitive: the larger the reflection, the more probable is the target detection. The larger is the noise, the more probable is the erroneous detection. 

Now, let us consider the quantum illumination scheme depicted in Fig. \ref{fig:1}(a). For such illumination, the following entangled state of two photons, equally distributed between $2d$ field modes, will be used:
\begin{equation}
\label{eqn:QI}
    |\Psi\rangle=\frac{1}{\sqrt{d}}\sum_{k=1}^d|1_k\rangle_{\mathrm{s}}|1_k\rangle_{\mathrm{i}}.
\end{equation}
Here, indices $\mathrm{s}$ and $\mathrm{i}$ indicate signal and idler modes, correspondingly; also,  $|1_k\rangle_{\mathrm{s}}$ and $|1_k\rangle_{\mathrm{i}}$ describe the states where there is just one photon in the signal and idler mode $k$ and no photons in the other modes. All signal modes are sent toward the target, while all idler modes are stored locally. When the reflected photon arrives, both the idler and signal modes are measured by projecting their state on the initial one (\ref{eqn:QI}).  Then, the probability to get a positive result for the target detection when the target is not actually present is $p_{\mathrm{false}}\approx n/d$. So, we have shown that the “false-positive” error probability can be hugely reduced for a wide-band (i.e., for $d\gg1$) field! At the same time, the probability of target detection when the object is really there is not changed that much $p_{\mathrm{true}}\approx n/d+R$ \cite{Lloyd1463}.

Unfortunately, the promise of QI scheme did not deliver equally powerful practical results. The history of the scheme development is well described and discussed in the recent work of Shapiro \cite{shapiro2020quantum}. First of all, the QI requires rather complicated non-local detection. Second, the mentioned huge and potentially unlimited advantage takes place only in comparison with single non-entangled photons. It appeared that by replacing single photons with unit-amplitude coherent states, one can perform not worse than with entangled ones \cite{shapiro2009quantum}. However, the scheme is ``saved'' when one goes beyond the single-photon approximation. It has been shown that implementing multi-photon entangled states (more precisely, a set of entangled modal pairs, each being in the squeezed vacuum state)
\begin{equation}
\label{eqn:multi-photon entangled}
    |\Psi\rangle=\sum_{n=0}^{\infty}\frac{\mu^\frac{n}2}{(\mu+1)^{\frac{n}2+1}}|n\rangle_{\mathrm{s}}|n\rangle_{\mathrm{i}},
\end{equation}
one can still get potentially 6 dB of SNR advantage in comparison with the optimal scheme implementing coherent states \cite{tan2008quantum, guha2009gaussian}.  For that purpose, one needs a rather involved  detection scheme; an example of such a scheme, based on the sum-frequency generation, has been recently suggested in \cite{zhuang2017optimum}. In Eq. (\ref{eqn:multi-photon entangled}), the vectors $|n\rangle_{\mathrm{s}}$ and $|n\rangle_{\mathrm{i}}$ describe $n$-photon Fock states in the signal and idler modes, respectively, $\mu$ is the average photon number in both modes.

The QI scheme was experimentally demonstrated at optical wavelengths \cite{lop,zhang2015entanglement}, and was shown to be quite robust with respect to  losses and noise \cite{zhang2015entanglement}.  Notice that for the optical wavelength region, the background noise is quite weak. So, thermal noise was artificially added to the signal. In 2015, a QI target-detection system that operates in the microwave regime was presented \cite{barz2015}. In this case, the target region can be interrogated at a microwave frequency, while the QI joint measurement, needed for target detection, is made at an optical frequency.  

A curious detail about the QI scheme is  that the condition for its functioning seems to negate the very prerequisite for its functioning. As such, a QI scheme requires entangled states. However, there is no more entanglement between the idler and signal, latter returning after the low-probability reflection and dilution with noise \cite{zhang2015entanglement}. A more resilient kind of quantum correlations, namely, quantum discord might be responsible for the QI gain \cite{korolkova2019quantum,weedbrook2016discord}.  
\begin{figure}[h]
    \centering
    \includegraphics[width=0.4\textwidth]{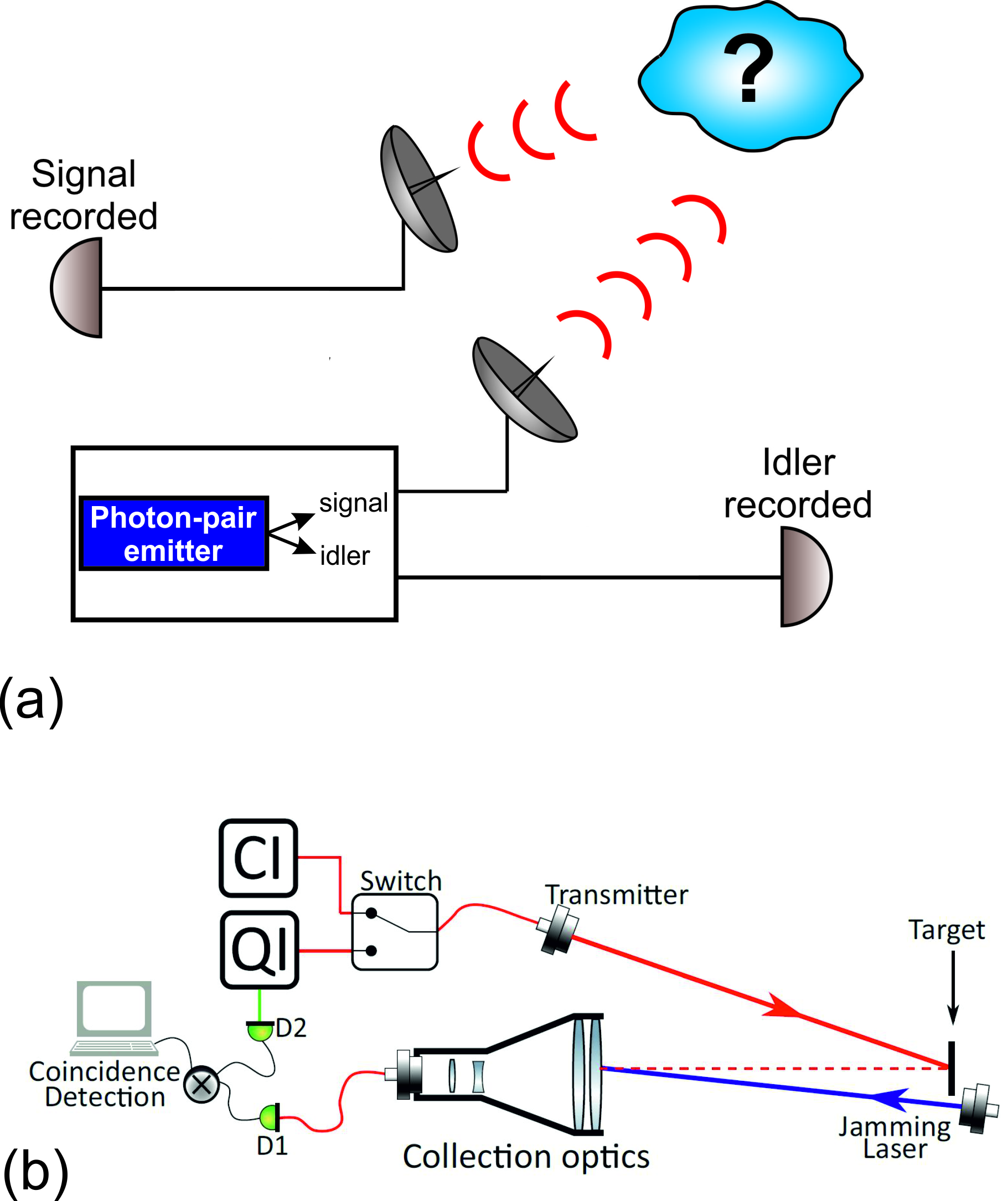}
    \caption{(a) A scheme of the microwave quantum-enhanced noise radar \cite{messaoudi2020quantum}. (b) A scheme of the optical diffusive target detection \cite{eng}. QI (CI) photons are the sources of quantum (classical) light. Detectors D1 and D2 are used to register the photons from the scattered signal and the idler; the jamming laser is used as background light source. Reprinted  with permission from \cite{eng}. Copyright (2019) by the American Physical Society."
 }
    \label{fig:2}
\end{figure}

\subsection{Beyond the original quantum illumination}

Notice that the original QI scheme seems quite limited in terms of practical realizations. Even without taking into account the difficulties with the projection-on-entangled-states measurement, the QI scheme assumed that the distance to the target is known and used to calculate the delay between the idler and the signal. Furthermore, a specific (thermal) noise model is implied, and the target is assumed to be just reflecting without introducing additional noise to the signal \cite{Lloyd1463,shapiro2020quantum}.  { So, a number of attempts were made to modify and enhance the QI. For example, it was suggested to simplify the detection of the returning signal and the idler making it local, to use conventional detection} schemes, and to avoid actual interference between the returning signal and idler fields (though, one will pay for it with a lower available improvement, for example, reducing possible gain to 3 dB \cite{guha2009gaussian}). 

Recently, a microwave scheme has appeared where one does not use the interference of the returning signal and the idler \cite{messaoudi2020quantum}. It is depicted in Fig. \ref{fig:2}. The entangled state (\ref{eqn:multi-photon entangled}) is produced in the superconducting JPA acting as a microwave parametric down-conversion source (see the section on the microwave photon-pair generation). In contrast to the scheme in Fig. \ref{fig:1}(b), the returning signal and the idler are recorded separately and the correlations of the recorded measurement data are analyzed \cite{messaoudi2020quantum}. A similar scheme was also implemented in the optical wavelength range for detecting a diffusely reflecting target \cite{eng}. 

Note that rough-surface target, inducing amplitude and phase noise, can also make getting quantum advantage by the QI scheme highly problematic (here we assume that the quantum advantage is the ability of a device to perform better, say, provide for a better SNR) with non-classical states in comparison with the classical ones, for the same spent resources (say, average number of photons of the signal field).  As was recently shown, a realistic so-called “Rayleigh-fading target”, inducing Rayleigh-distributed amplitudes and uniformly distributed phases of the scattered field, completely negates the possibility to obtain the 3 dB gain achievable for the local detection. Also, it reduces to 3 dB possible gain from the advanced detection scheme of \cite{zhuang2017optimum}, which ideally allows 6 dB of SNR gain for the QI scheme with a noiseless reflective target \cite{zhuang2017quantum}.  

However, even if the target completely suppresses the spatial correlations of the illuminating field, one can still resort to the temporal correlations between the emitted photons \cite{eng}. In that case, the registration of non-classicality can be done just by the standard photo-counting measuring coincidences using a pair of detectors. In Fig. \ref{fig:2}(b), the scheme of the experiment carried out in \cite{eng} is shown. { This scheme is sufficiently simple and illustrative, so, let us show how the quantum advantage was sought for in this case.} In the classical case of a coherent illuminating source, in the absence of background, the probability of detecting a signal photon in a given time-bin is $\epsilon P_{\mathrm{s}}$. Here $P_{\mathrm{s}}$ is the probability of scattering toward the detector, and $\epsilon$ is the overall collection efficiency, which incorporates all the collection losses and the detector efficiency. In the presence of background, the probability of detecting a background photon in the same time-bin is $P_{\mathrm{b}}$. The SNR is therefore simply given by $SNR_{\mathrm{classic}}=\epsilon P_{\mathrm{s}}/P_{\mathrm{b}}$. In the quantum case and in the absence of background, the probability of simultaneous detection of the signal and idler photons is $P_{\mathrm{si}}$. In the presence of background, the probability of accidentally detecting a background photon in coincidence with an idler photon is $P_\mathrm{b} P_{\mathrm{i}}$, where $P_{\mathrm{i}}$ is the probability of detecting an idler photon. The SNR is therefore given by $SNR_{\mathrm{quant}}=\epsilon P_{\mathrm{si}}/P_{\mathrm{i}}P_\mathrm{b}$.  
So, the ratio of the quantum and classical SNRs does not include the probability of the background photon detection. { For the squeezed vacuum illumination field (\ref{eqn:multi-photon entangled}), it  is just proportional to the inverse average number of photons $\mu$,  provided that $\mu\ll1$}
\begin{equation}
    \frac{SNR_{\mathrm{quant}}}{SNR_{\mathrm{classic}}}=\frac{P_{\mathrm{si}}}{P_{\mathrm{i}}P_{\mathrm{s}}}\sim \frac{1}{\mu}.
\end{equation}
   { The work  \cite{eng} claims that  the quantum enhancement on the order of $10^3$ was demonstrated in this way}. Alas, the  classical detection taken for comparison is far from being the optimal one for the case. { Very recently, it has been shown that by implementing thermal fields of various temperatures, correlated just by asymmetric beam-splitting, one can reduce the quantum advantage to already mentioned 3dB \cite{Jonsson2020ACB}. }
 Besides, lately the temporal correlations of twin-photons have been suggested for enhancing the performance of a lidar \cite{liu2019enhancing},  and a scheme utilizing simultaneously both spatial and temporal correlations has been proposed in \cite{zhang2020multidimensional}. 

{ One should emphasize that very recently rather serious doubts were voiced about the very existence  of quantum advantage in the so far realized QI radar schemes \cite{shapiro2020quantum,Cho1556,Jonsson2020ACB,p1,p2}.} Also, practical perspectives and cost of a presumably lab-successful QI radar were also predicted to be quite pessimistic  \cite{9114614}.

\subsection{Quantum ranging}
As noted above, { even for the possibility to hope gaining} an advantage with QI, one needs to know the distance to the target. On the other hand, it has been recently shown that one can perform ranging of the target with a quantum advantage \cite{maccone2020quantum}.The scheme of this quantum radar is depicted in Fig. \ref{fig:1}(b). Let us explain it with our standard testing of the state of entangled photon pairs. Both photons impinge on a point target, and then the scattered photons  are registered by two detectors at some distance one from another. These detectors measure the second-order delayed intensity correlation function of the scattered field. Then, one can get $\sqrt{2}$ fold advantage in the standard deviation of each average coordinate of the target. The advantage is achieved with the following state
\begin{equation}
\label{eqn:phi_n}
    |\phi\rangle_n\sim\sum\limits_j \Bar{\phi}_j|n_j\rangle,
\end{equation}
where the function $\Bar{\phi}_j$  describes how photons are distributed through the modes, and $|n_j\rangle$ denotes the state with $n$ photons in the $j$-th mode and vacuum in all other modes. For the case of entangled photon pairs, $n=2$, and by measuring the second-order intensity correlation function, the following result is obtained \cite{maccone2020quantum}:
\begin{equation}
\label{eqn:G2}
    G^{(2)}(\vec r_1,t_1,\vec r_2,t_2)\sim |f(2[\Tilde{t}-t_0],2[\vec r-\vec r_0])|^2,
\end{equation}
where the function $f(t,\vec r)$ is the Fourier-transform of the wave-function $\Bar{\phi}$;  $\Tilde{t}=(t_1+t_2)/2$, and $t_j$, $j=1,2$ is the photon arrival time at the first or the second detector; $\vec r=(\vec r_1+\vec r_2)/2$, and $\vec r_j$ is the position of the $j$-th detector; $t_0$ is the state generation time, $\vec r_0$ is the target position.

For two independent photons with the state described by the product of the wavefunctions  $|\phi\rangle_1$ (\ref{eqn:phi_n}), the second-order intensity correlation function $G^{(2)}(\vec r_1,t_1,\vec r_2,t_2)$ (\ref{eqn:G2}) is the product of two independently measured first-order correlation functions (i.e., intensities)
\begin{equation}
\label{eqn:G1}
    G^{(1)}(\vec r_1,t_1)\sim |f(t_j-t_0,\vec r_j-\vec r_0)|^2,
\end{equation}
So, the comparison of Eqs. (\ref{eqn:G2}) and (\ref{eqn:G1}) shows that the entangled state (\ref{eqn:phi_n}) for $n=2$ indeed gives $\sqrt{2}$ reduction in the standard deviation of the average time of photon arrival in comparison with the case of two unentangled photons, and the same advantage in standard deviation for each of the two components of the target average position.
\begin{figure}
    \centering
    \includegraphics[width=0.4\textwidth]{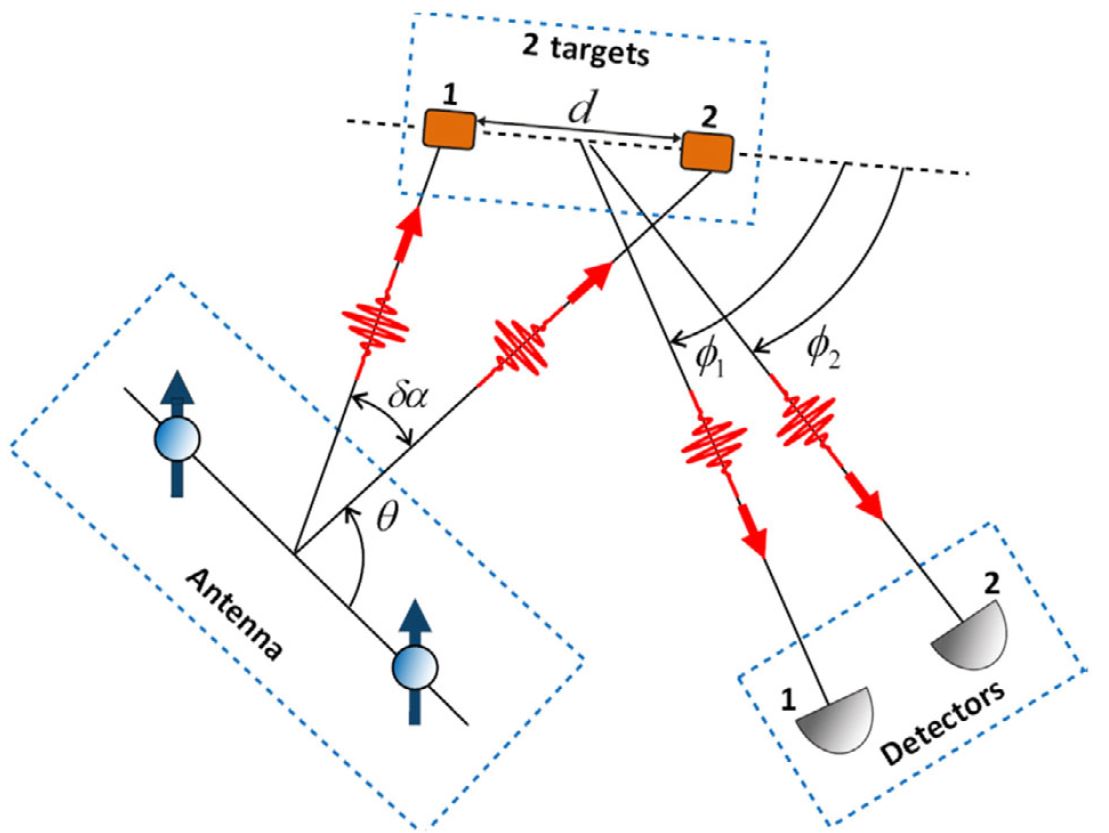}
    \caption{A scheme of a target-resolving quantum radar with spatially and temporally correlated photon pairs emitted by the simplest two-emitter quantum antenna \cite{peshko2019quantum}.}
    \label{fig:4}
\end{figure}
However, the practical realization of this advantage seems rather problematic. To find the maximum of the function (\ref{eqn:G2}), it is necessary to probe all the detector positions for all possible arrival times. This problem can be alleviated to some extent by considering the non-ideally entangled state (\ref{eqn:phi_n}), but rather a partially entangled state with finite correlation time and length, at a cost of reducing the advantage \cite{maccone2020quantum}. Losses and noise also pose a quite hard problem in the described scheme, and lead to further degradation of the gain. 
On the other hand, this scheme holds a considerable promise as well. Indeed, for the state of $n$ entangled photons (\ref{eqn:phi_n}), the scheme can potentially give $\sqrt{n}$ gain for each target coordinate.

\subsection{Quantum radar for target imaging}
The final task for the quantum radar/lidar considered here is the target recognition, i.e., the determination of the target spatial shape.  In its simplest version, this problem can be reduced to the one typical for imaging: estimating the distance between two close point targets. Concerning this task, the main obstacle is the same as for the general imaging problem: the field diffraction. A point object is imaged not into a point, but into some region/spot. For two or more point objects, these regions can be strongly overlapping and point objects cannot be resolved in the measurement result.  For classical radar imaging, the fundamental resolution is dictated by the Rayleigh limit \cite{rayleigh1879xxxi}. So, the task for the quantum radar is to provide possibility to go beyond this limit. 

Here, we will discuss a generic imaging-like scheme (like that in Fig. \ref{fig:1}(b)). Like for the quantum ranging scheme, the entire generated field is directed toward the target, and the measurement set-up infers the correlations between the returning photons.  Recently, a simple realization of this scheme has been suggested \cite{peshko2019quantum}. In this set-up (dubbed the ‘quantum noise radar’), a photon pair, produced by a simple antenna of two interacting two-level emitters, irradiates the target composed of several point objects (see Fig. \ref{fig:4}). Then, the delayed second-order intensity correlation function of the returning field is measured. It was demonstrated that using entangled photons one can vastly outperform the same measurement with two uncorrelated photons \cite{peshko2019quantum}. More precisely, it is possible to avoid the so-called “Rayleigh’s Curse” using the correlations of the registered photons. The “Rayleigh’s Curse” is a recently appeared sobriquet for the well-known diffraction-caused loss of ability to resolve two close point sources by common intensity measurements \cite{tsang2016quantum}. Formally, this “curse” appears as the information (\ref{eqn:fisher}) about the distance between the objects, calculated for such intensity measurements, tending to zero, when the distance tends to zero. However, the quantum analogy of the Fisher information (\ref{eqn:fisher}) is not zero in the limit of zero distance between the objects. It means that one can find a measurement beat frequency in the diffraction limit, thus “dispelling” the “curse”.
The scheme, suggested in \cite{peshko2019quantum}, indeed allows one to dispel the “curse”. The information stays finite for an arbitrary small distance between the objects providing their resolution within finite measurement time. Moreover, for inferring the distances between more than two objects, the “quantum noise radar” also allows one to outperform sensing with the uncorrelated photons. 

Here, one should notice that there are possibilities to significantly enhance the resolution in the target recognition problem even using classical illuminating fields by designing sophisticated “quantum-inspired” detection schemes, for example, for building telescopes for astronomical applications  \cite{tsang2019resolving}.

\subsection{Single-photon lidars}
Devices, designed specifically for quantum detection, can also greatly enhance the performance of far-field sensing schemes, even if classical illuminating fields are used. For example, such a typically quantum device as an array of single-photon avalanche diodes (SPAD), suitable for registering non-classical photon statistics, can bring a considerable advantage to lidar schemes \cite{kirmani2014first, shin2016photon}. 
\begin{figure}
    \centering
    \includegraphics[width=0.45\textwidth]{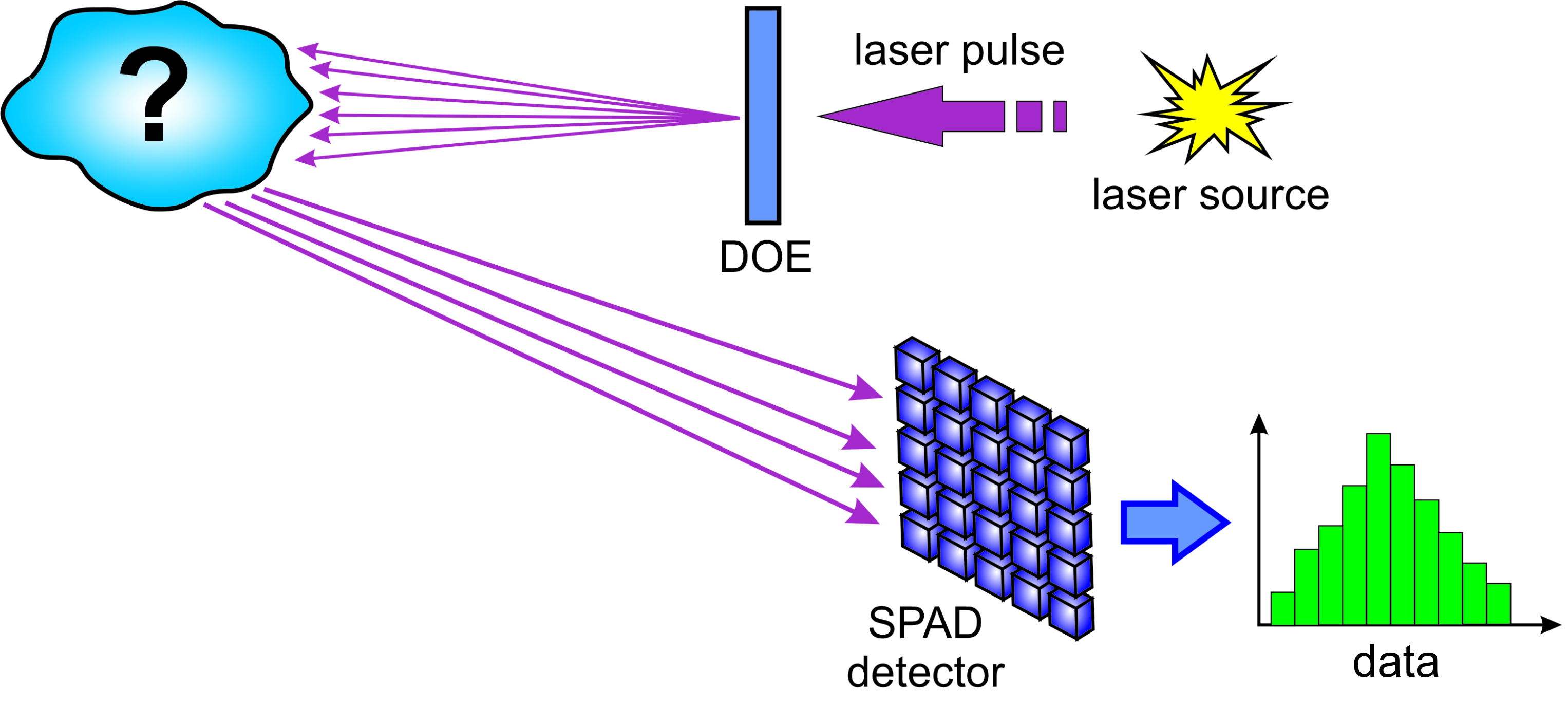}
    \caption{A scheme of the lidar with a single-photon detector array: initial laser beam is split into many beamlets using a diffractive optical element (DOE), the returning signal is imaged with the SPAD.    }
    \label{fig:5}
\end{figure}
A scheme of the implementation of a single-photon lidar with a detector array is depicted in Fig. \ref{fig:5}. A laser pulse is split into a large number of narrow beams propagating toward the target. Then, an array of single-photon detectors registers the returning signal, which is at the single-photon level. After collecting sufficient data statistics, the object characteristics can be inferred. 
Such single-photon lidars are already commercially available and widely used. They provide a number of important advantages, such as high-sensitivity, possibility of using weaker laser sources, eye-safety, low cost, potentially higher area coverage performance, improved depth resolution, and accuracy. These advantages come at a price: necessity of high computing power and advanced methods for object reconstruction and noise removal, especially if one aims at real-time imaging of moving targets. Curiously, in the spirit of the “quantum-inspired imaging” \cite{tsang2019resolving}, it has been very recently claimed that single-photon lidars are able to achieve twofold enhancement of spatial resolution over the diffraction limit for distances of more than 8 km \cite{li2020super}. 

\section{Generation of entangled photon pairs}
In all our quantum radar examples, we have considered twin-photon generators as a basic tool of entangled states for realizing quantum advantage. Here, we give a brief description of the ways to build these generators aiming at the radar applications. Generally, they are based on the second- and the third-order nonlinearities. However, particular implementations significantly depend on the targeted frequency range. Also, here we address such emerging field as quantum antennas. 

\subsection{Twin-photons for optical frequencies}
\begin{figure}[h]
    \centering
    \includegraphics[width=0.5\textwidth]{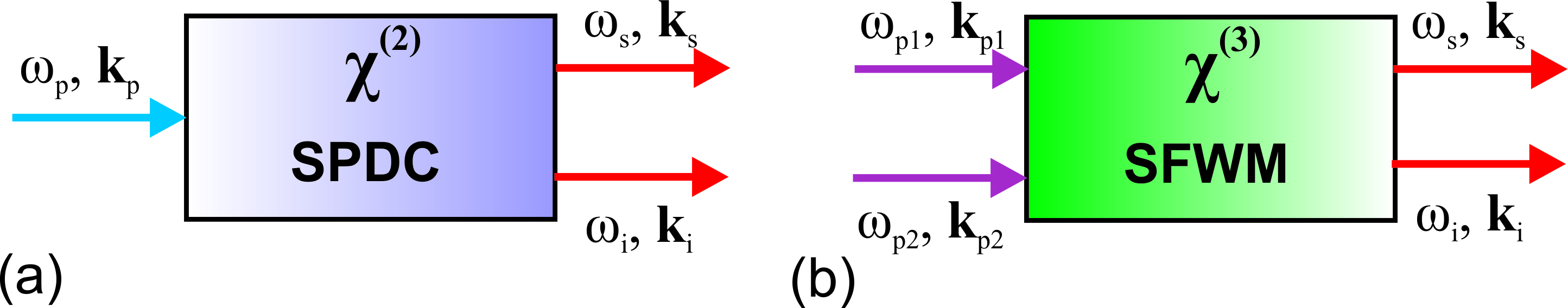}
    \caption{The schemes of the (a) SPDC and (b) SFWM sources of entangled photon pairs.}
    \label{fig:6}
\end{figure}

For optical frequencies, the cheapest and most widely used mechanisms of entangled states generation are those based on spontaneous parametric-down conversion (SPDC) and spontaneous four-wave mixing (SFWM) \cite{drummond2013}. The illustration of their operating principles are shown in Fig. \ref{fig:6}. For SPDC, the second-order nonlinear medium is pumped by a pump beam at some frequency $\omega_{\mathrm{p}}$.

Each pumping photons can be split giving birth to two photons with lower frequencies $\omega_{\mathrm{s}}$ and $\omega_{\mathrm{i}}$, named “signal” and “idler”, satisfying energy conservation, $\omega_{\mathrm{s}}+\omega_{\mathrm{i}}=\omega_{\mathrm{p}}$. Momentum conservation requires that $\vec k_{\mathrm{s}}+\vec k_{\mathrm{i}}=\vec k_{\mathrm{p}}$, where $\vec k_{\mathrm{s,i,p}}$ are the wavevectors for the signal, idler, and pump photons, respectively. These conservation laws lead to the appearance of entanglement between the signal and idler photons. Indeed, registering a photon flying in the direction  $\vec k_{\mathrm{s}}$ means that there is another photon flying in the direction $\vec k_{\mathrm{p}}-\vec k_{\mathrm{s}}$. For a weak pump, a superposition of the entangled state (\ref{eqn:QI}) with the vacuum can be produced.
The SPDC-based sources can be routinely produced now, for example, using potassium titanyl phosphate crystals, or similar media exhibiting the second-order nonlinearity \cite{drummond2013}. 

In the SFWM process occurring by the third-order nonlinearity, two pump photons are converted into signal and idler photons while satisfying the energy and momentum conservation laws providing the entanglement of the generated state (which is also of the form (\ref{eqn:QI}) for a weak pump). Photon pairs can be produced by the SFWM mechanism even in such common structures as index-guiding optical fibers; glass usually possesses some weak third-order nonlinearity \cite{drummond2013, solntsev2017path}.

\subsection{Entangled photons for microwave frequencies}
\begin{figure}[h]
    \centering
    \includegraphics[width=0.45\textwidth]{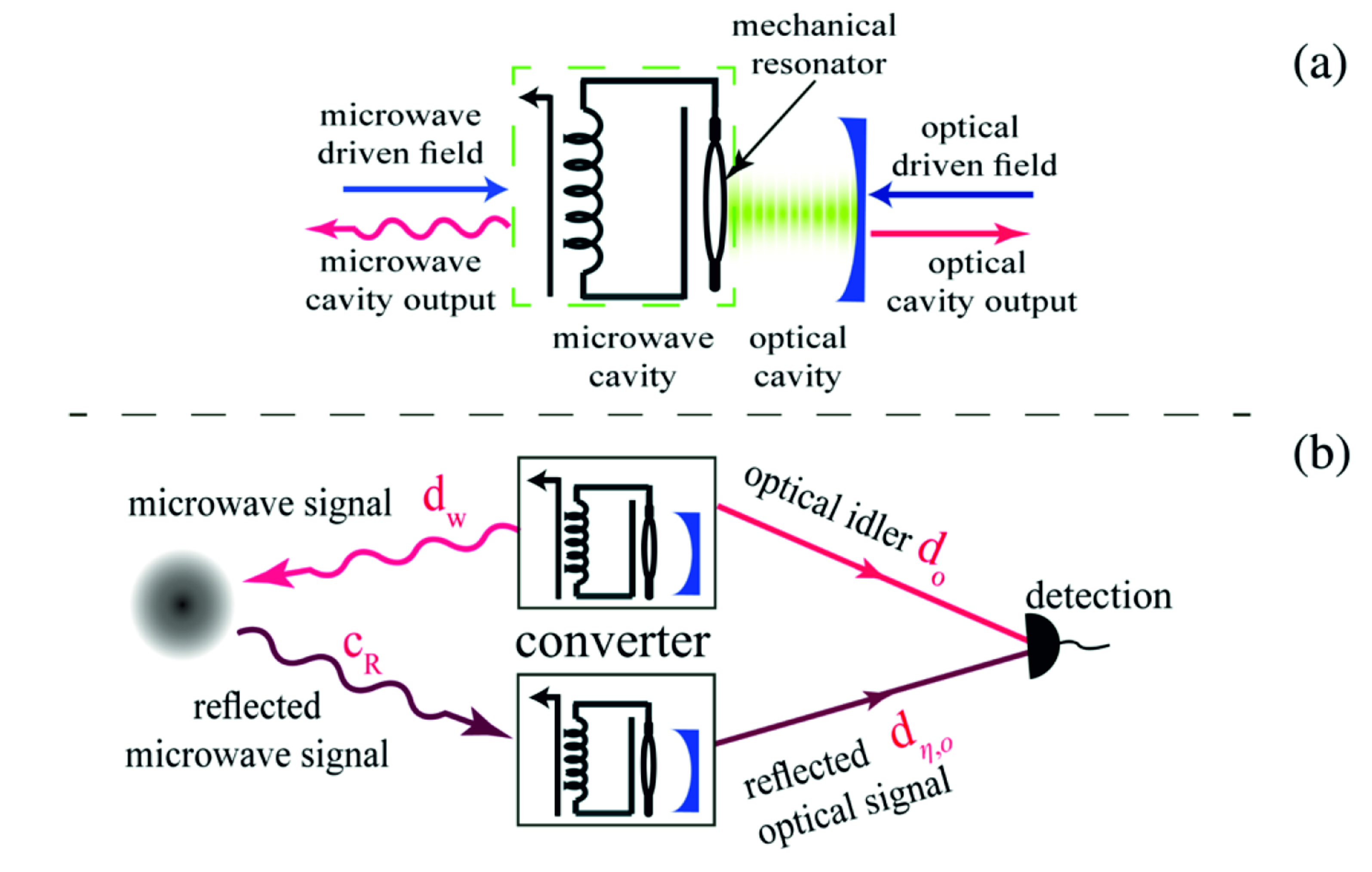}
    \caption{(a) The scheme of the converter device in which a mechanical resonator couples  microwave and optical cavities. (b) The scheme of the microwave-optical QI using converters shown in the panel (a). The transmitter's converter entangles signal optical and microwave fields. The receiver's converter transfers the returning field back to the optical domain. Reprinted  with permission from Ref. \cite{barz2015}. Copyright (2015) by the American Physical Society.}
    \label{fig:7}
\end{figure}

Notice that creating entangled fields in the microwave frequency region is considerably more difficult than at the optical wavelengths. There are no readily available nonlinearities. Nowadays, the quantum radar researchers exploit mostly two radically different ways to create quantum-correlated microwave fields for quantum radar schemes. The first is to create initially correlated field in the optical wavelengths range, and then to transfer the entanglement into the microwave wavelength field. The second way is to use directly the available (and rather expensive) methods, such as microwave parametric amplifiers (based on superconducting Josephson junctions \cite{makhlin2001quantum}) to generate entangled microwave fields. 
The first way was realized in the experiment on the microwave realization of the QI scheme \cite{barz2015}. This setup is based on the coupling between microwave and optical cavities, which allows one to transfer the entanglement creating quantum correlations between the signal illuminating the target and the optical idler (Fig. \ref{fig:7}(a)). The optical signal modulates the microwave cavity and entangles the microwave signal with the stored idler. Then the microwave signal travels to the target, returns, and gets quantum correlated with the optical mode. Finally, the correlation measurement of this optical mode and the stored idler is performed (Fig. \ref{fig:7}(b)). 

The second way is, essentially, a microwave analog of the optical SPDC (Fig.~\ref{fig:5}(a)) realized by the JPA. Such a parametric amplifier is able to generate entangled states of the squeezed vacuum type (\ref{eqn:multi-photon entangled}). For example, the JPA device discussed for the purpose of the entangled state generation in the recent work \cite{zhong2013squeezing}, is realized as a transmission line resonator with the resonant frequency modulated with the help of the superconducting loop interrupted by two Josephson junctions, and requires millikelvin temperatures for operations.  
Recently, JPA state generators have been successfully used for the implementation of purely microwave QI-like schemes  \cite{luong2020, barzanjeh2020microwave}.
\begin{figure}[h]
    \centering
    \includegraphics[width=0.25\textwidth]{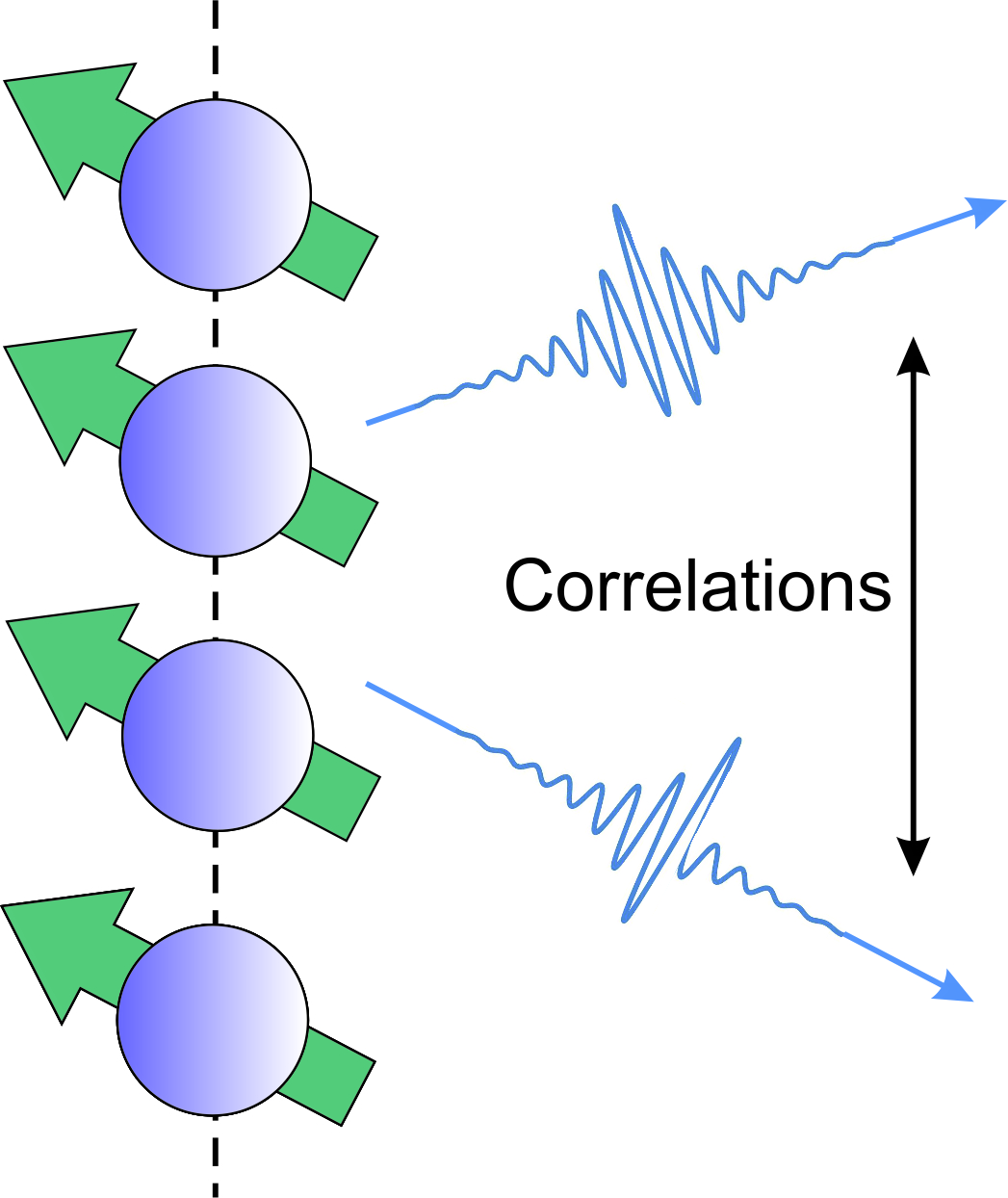}
    \caption{The  scheme of 1D array antenna of identical quantum dipoles for generating entangled photon pairs \cite{mikhalychev2018synthesis}.}
    \label{fig:9}
\end{figure}

\subsection{Quantum antennas}
Finally, we mention another way to generate propagating entangled states potentially suitable for the quantum radar and drastically different from the photon mixing schemes described above: quantum antenna arrays. They comprise sets of quantum emitters arranged in the manner typical for conventional classical arrays. The purpose of such antennas is to use both quantum correlations of the emitters and classical interference of the emitted states to shape the correlations and spatiotemporal characteristics of the emitted field in the far-field zone \cite{ourreview}. An example of such a quantum antenna as a source for a target-recognizing quantum radar was considered in the previous section \cite{peshko2019quantum}. There, just two interacting two-level emitters were shown to be sufficient to create photon pairs achieving super-resolved target imaging. 
Generally, manipulating the state of the multi-emitter antenna even with the simplest geometry (such, for example, as a 1D array considered in \cite{mikhalychev2018synthesis}, Fig. \ref{fig:9}), one can strongly change the correlation shape. For example, one can make emitted photons fly together or in opposite directions, or completely suppress the field in the far-field zone \cite{mikhalychev2018synthesis}.  Quantum antennas can be realized in a number of different ways: with quantum dots, atoms and ions in traps, fluorescent nanodiamonds, etc. They can also be implemented by spatially arranging elements performing three or/and four-wave mixing \cite{ourreview,Marino:19}.

\section{Conclusion and outlook}
 { In this article, we have reviewed the basic theory, design, preliminary experiments, and potential applications of quantum radars and lidars.  Given the high potential of quantum technologies for this branch of science and engineering, we find it crucial to attract the traditional radar community to this area of research and to leverage the conventional radar concepts to improve the design of radars { and lidars} based on quantum technologies. Here, we have also discussed how traditional quantum optical concepts (such as entanglement) can be and have been translated and applied to the microwave and THz frequency ranges despite the drastically different implementations in optics.   We have described several ideas for enhancing SNR by quantum effects for the target detection, ranging and imaging, referenced several experiments (both in optics \cite{zhang2015entanglement,lop, eng, liu2019enhancing} and microwaves \cite{luong2020, barz2015, luong2020quantum})  attempting to overcome classical limits with the QI quantum radar protocol (very recently, an excellent popular introduction to QI radars and lidars has appeared \cite{french}). For the moment, there is an opinion that in the {microwave} region quantum advantage still seems out of reach. Also, technological difficulties remain formidable, and they are considerably limiting perspectives of practical realizations  \cite{9114614}. In the same time,  there are schemes potentially allowing one to go beyond QI \cite{cohen2019thresholded},  outlining the way to achieve quantum advantages in ranging and target imaging employing entangled photon pairs \cite{maccone2020quantum,peshko2019quantum}.  In particular, quantum radars {and lidars} open the way to go beyond the classical resolution limit for target imaging \cite{french}.
 
We expect that the ideas, discussed in this article, will be increasingly appreciated and actively employed by the quantum optical and electrical engineering communities. We also expect that in the coming years the usage of quantum technological developments in far-field sensing will be steadily increasing in the form of novel efficient detecting schemes, and also novel target sensing, ranging and imaging protocols}. 

\begin{acknowledgments} 
{G.Y.S. and A.B. acknowledge support from the H2020, project TERASSE 823878, D. M. and S.V. acknowledge
support from the EU Flagship on Quantum Technologies,
project PhoG 820365.}
\end{acknowledgments}



\begin{thebibliography}{99}
\providecommand{\url}[1]{#1}
\csname url@samestyle\endcsname
\providecommand{\newblock}{\relax}
\providecommand{\bibinfo}[2]{#2}
\providecommand{\BIBentrySTDinterwordspacing}{\spaceskip=0pt\relax}
\providecommand{\BIBentryALTinterwordstretchfactor}{4}
\providecommand{\BIBentryALTinterwordspacing}{\spaceskip=\fontdimen2\font plus
\BIBentryALTinterwordstretchfactor\fontdimen3\font minus
  \fontdimen4\font\relax}
\providecommand{\BIBforeignlanguage}[2]{{%
\expandafter\ifx\csname l@#1\endcsname\relax
\typeout{** WARNING: IEEEtran.bst: No hyphenation pattern has been}%
\typeout{** loaded for the language `#1'. Using the pattern for}%
\typeout{** the default language instead.}%
\else
\language=\csname l@#1\endcsname
\fi
#2}}
\providecommand{\BIBdecl}{\relax}
\BIBdecl

\bibitem{einstein1935}
A.~Einstein, B.~Podolsky, and N.~Rosen, ``Can quantum-mechanical description of
  physical reality be considered complete?'' \emph{Phys. Rev.}, vol.~47, pp.
  777--780, 1935.

\bibitem{nielsen_chuang_2010}
M.~A. Nielsen and I.~L. Chuang, \emph{Quantum Computation and Quantum
  Information: 10th Anniversary Edition}.\hskip 1em plus 0.5em minus
  0.4em\relax Cambridge University Press, 2010.

\bibitem{dowling2015}
J.~P. {Dowling} and K.~P. {Seshadreesan}, ``Quantum optical technologies for
  metrology, sensing, and imaging,'' \emph{J. Light. Technol.}, vol.~33,
  no.~12, pp. 2359--2370, 2015.

\bibitem{Haroche2020FromCT}
S.~Haroche, M.~Brune, and J.-M. Raimond, ``From cavity to circuit quantum
  electrodynamics,'' \emph{Nat. Phys.}, vol.~16, pp. 243--246, 2020.

\bibitem{schuster}
D.~Schuster, ``Hybrid quantum systems with circuit quantum electrodynamics,''
  \emph{APS}, vol. 2012, p. 007, 2012.

\bibitem{blais}
A.~Blais, S.~Girvin, and W.~Oliver, ``Quantum information processing and
  quantum optics with circuit quantum electrodynamics,'' \emph{Nat. Phys.},
  vol.~16, pp. 1--10, 2020.

\bibitem{carusotto}
I.~Carusotto, A.~Houck, A.~Kollár, P.~Roushan, D.~Schuster, and J.~Simon,
  ``Photonic materials in circuit quantum electrodynamics,'' \emph{Nat. Phys.},
  vol.~16, pp. 1--12, 2020.

\bibitem{ourreview}
G.~Y. Slepyan, S.~Vlasenko, and D.~Mogilevtsev, ``Quantum antennas,''
  \emph{Adv. Quant. Tech.}, vol.~3, no.~4, p. 1900120, 2020.

\bibitem{Rich}
M.~A. Richards, J.~A. Scheer, and W.~A. Holm, ``Principles of modern radar:
  Basic principles,'' 2013.

\bibitem{1880MNRAS..40..254R}
L.~Rayleigh, ``Investigations in optics, with special reference to the
  spectroscope,'' \emph{MNRAS}, vol.~40, p. 254, 1880.

\bibitem{Abbe2009}
E.~Abbe, ``Beitr{\"a}ge zur theorie des mikroskops und der mikroskopischen
  wahrnehmung,'' \emph{Archiv f{\"u}r mikroskopische Anatomie}, vol.~9, pp. 413
  -- 468, 2009.

\bibitem{Helstrom:70}
C.~W. Helstrom, ``Resolvability of objects from the standpoint of statistical
  parameter estimation,'' \emph{J. Opt. Soc. Am.}, vol.~60, no.~5, pp.
  659--666, 1970.
  

\bibitem{tsang2015} M.~Tsang, "Quantum limits to optical point-source localization," Optica \emph{2}, 646-653 (2015).

 \bibitem{tsang2019resolving}
M.~Tsang, ``Resolving starlight: a quantum perspective,'' \emph{Contemp.
  Phys.}, vol.~60, no.~4, pp. 279--298, 2019.
 

\bibitem{784050}
{L. Guosui}, {G. Hong}, and {S. Weimin}, ``Development of random signal
  radars,'' \emph{IEEE Trans. Aerosp. Electron. Syst.}, vol.~35, no.~3, pp.
  770--777, 1999.


\bibitem{Lukin_2001}
K.~A. {Lukin}, ``Noise radar technology,'' \emph{Telecommunications and Radio
  Engineering}, vol.~55, no.~12, 2001.

\bibitem{Thayaparan2006NoiseRT}
T.~Thayaparan and C.~Wernik, ``Noise radar technology basics: Defence R{\&}D
  Canada/Ottawa, Technical memorandum,  DRDC Ottawa TM  2006-266,'' 2006.

\bibitem{giovannetti2011advances}
V.~Giovannetti, S.~Lloyd, and L.~Maccone, ``Advances in quantum metrology,''
  \emph{Nat. Phot.}, vol.~5, no.~4, p. 222, 2011.

\bibitem{Lloyd1463}
S.~Lloyd, ``Enhanced sensitivity of photodetection via quantum illumination,''
  \emph{Science}, vol. 321, no. 5895, pp. 1463--1465, 2008.

\bibitem{dowling}
J.~P. Dowling, ``Quantum optical metrology--the lowdown on high-{N00N}
  states,'' \emph{Contemp. Phys.}, vol.~49, no.~2, pp. 125--143, 2008.

\bibitem{lanza}
M.~Lanzagorta, \emph{Quantum Radar}, 2011, vol.~3.

\bibitem{simon}
D.~Simon, G.~Jaeger, and A.~Sergienko, \emph{Quantum Metrology, Imaging, and
  Communication}, 2016.
  
\bibitem{zhang2015entanglement}
Z.~Zhang, S.~Mouradian, F.~N. Wong, and J.~H. Shapiro, ``Entanglement-enhanced
  sensing in a lossy and noisy environment,'' \emph{Phys. Rev. Lett.}, vol.
  114, no.~11, p. 110506, 2015.

\bibitem{lop}
E.~Lopaeva, I.~Ruo~Berchera, I.~Degiovanni, S.~Olivares, and G.~Brida,
  ``Experimental realization of quantum illumination,'' \emph{Phys. Rev.
  Lett.}, vol. 110, p. 153603, 2013.

\bibitem{eng}
D.~G. England, B.~Balaji, and B.~J. Sussman, ``Quantum-enhanced standoff
  detection using correlated photon pairs,'' \emph{Phys. Rev. A}, vol.~99, p.
  023828, 2019.

\bibitem{luong2020}
D.~{Luong}, C.~W.~S. {Chang}, A.~M. {Vadiraj}, A.~{Damini}, C.~M. {Wilson}, and
  B.~{Balaji}, ``Receiver operating characteristics for a prototype quantum
  two-mode squeezing radar,'' \emph{IEEE Trans. Aerosp. Electron. Syst.},
  vol.~56, no.~3, pp. 2041--2060, 2020.


\bibitem{liu2019enhancing}
H.~Liu, D.~Giovannini, H.~He, D.~England, B.~J. Sussman, B.~Balaji, and A.~S.
  Helmy, ``Enhancing lidar performance metrics using continuous-wave
  photon-pair sources,'' \emph{Optica}, vol.~6, no.~10, pp. 1349--1355, 2019.

\bibitem{zhang2020multidimensional}
Y.~Zhang, D.~England, A.~Nomerotski, P.~Svihra, S.~Ferrante, P.~Hockett, and
  B.~Sussman, ``Multidimensional quantum-enhanced target detection via
  spectrotemporal-correlation measurements,'' \emph{Phys. Rev. A}, vol. 101,
  no.~5, p. 053808, 2020.

\bibitem{barz2015}
S.~Barzanjeh, S.~Guha, C.~Weedbrook, D.~Vitali, J.~H. Shapiro, and
  S.~Pirandola, ``Microwave quantum illumination,'' \emph{Phys. Rev. Lett.},
  vol. 114, p. 080503, 2015.

\bibitem{shapiro2020quantum}
J.~H. Shapiro, ``The quantum illumination story,'' \emph{IEEE Trans. Aerosp.
  Electron. Syst.}, vol.~35, no.~4, pp. 8--20, 2020.

\bibitem{Cho1556}
A.~Cho, ``The short, strange life of quantum radar,'' vol. 369, no. 6511, pp.
  1556--1557, 2020.

\bibitem{Jonsson2020ACB}
R. Jonsson, R. Di Candia, M. Ankel, A. Strom and G. Johansson, "A comparison between quantum and classical noise radar sources," 2020 IEEE Radar Conference (RadarConf20), Florence, Italy, 2020, pp. 1-6.

\bibitem{p1} N. Messaoudi, C. W. S. Chang, A. M. Vadiraj, C. M. Wilson, J. Bourassa and B. Balaji, "Practical Advantage in Microwave Quantum Illumination," 2020 IEEE Radar Conference, Florence, Italy, 2020, pp. 1-5.

\bibitem{p2} S. Barzanjeh, S. Pirandola, D. Vitali and J. Fink, "Microwave quantum illumination with a digital phase-conjugated receiver," 2020 IEEE Radar Conference (RadarConf20), Florence, Italy, 2020, pp. 1-6.



\bibitem{9114614}
F.~{Daum}, ``A system engineering perspective on quantum radar,'' in \emph{2020
  IEEE International Radar Conference (RADAR)}, 2020, pp. 958--963.

\bibitem{cohen2019thresholded}
L.~Cohen, E.~S. Matekole, Y.~Sher, D.~Istrati, H.~S. Eisenberg, and J.~P.
  Dowling, ``Thresholded quantum lidar: Exploiting photon-number-resolving
  detection,'' \emph{Phys. Rev. Lett.}, vol. 123, no.~20, p. 203601, 2019.

\bibitem{maccone2020quantum}
L.~Maccone and C.~Ren, ``Quantum radar,'' \emph{Phys. Rev. Lett.}, vol. 124,
  no.~20, p. 200503, 2020.

\bibitem{kirmani2014first}
A.~Kirmani, D.~Venkatraman, D.~Shin, A.~Cola{\c{c}}o, F.~N. Wong, J.~H.
  Shapiro, and V.~K. Goyal, ``First-photon imaging,'' \emph{Science}, vol. 343,
  no. 6166, pp. 58--61, 2014.

\bibitem{shin2016photon}
D.~Shin, F.~Xu, D.~Venkatraman, R.~Lussana, F.~Villa, F.~Zappa, V.~K. Goyal,
  F.~N. Wong, and J.~H. Shapiro, ``Photon-efficient imaging with a
  single-photon camera,'' \emph{Nat. Comm.}, vol.~7, no.~1, pp. 1--8, 2016.

\bibitem{PhysRevLett.96.230401}
Z.~Hradil, D.~Mogilevtsev, and J.~\ifmmode \check{R}\else
  \v{R}\fi{}eh\'a\ifmmode~\check{c}\else \v{c}\fi{}ek, ``Biased tomography
  schemes: An objective approach,'' \emph{Phys. Rev. Lett.}, vol.~96, p.
  230401, 2006.


\bibitem{kono} Kono, S., Koshino, K., Tabuchi, Y., et al.\ 2018, Nature Physics, 14, 546. doi:10.1038/s41567-018-0066-3

\bibitem{micro} R. Lescanne, S. Deleglise, E. Albertinale, U. Reglade, Th. Capelle, E. Ivanov, Th. Jacqmin, Z. Leghtas, and E. Flurin, Irreversible Qubit-Photon Coupling for the Detection of Itinerant Microwave Photons
Phys. Rev. X 10, 021038 (2020). 

\bibitem{PhysRevLett.123.231107}
M.~Tse and et~al., ``Quantum-enhanced advanced ligo detectors in the era of
  gravitational-wave astronomy,'' \emph{Phys. Rev. Lett.}, vol. 123, p. 231107,
  2019.

\bibitem{PhysRevLett.59.2044}
C.~K. Hong, Z.~Y. Ou, and L.~Mandel, ``Measurement of subpicosecond time
  intervals between two photons by interference,'' \emph{Phys. Rev. Lett.},
  vol.~59, pp. 2044--2046, 1987.

\bibitem{PhysRevLett.23.880}
J.~F. Clauser, M.~A. Horne, A.~Shimony, and R.~A. Holt, ``Proposed experiment
  to test local hidden-variable theories,'' \emph{Phys. Rev. Lett.}, vol.~23,
  pp. 880--884, 1969.

\bibitem{9096523}
G.~W. {Hanson}, ``Aspects of quantum electrodynamics compared to the classical
  case: Similarity and disparity of quantum and classical electromagnetics,''
  \emph{IEEE Antennas and Propagation Magazine}, vol.~62, no.~4, pp. 16--26,
  2020.

\bibitem{doi:10.1116/1.5126696}
K.~C. Tan and H.~Jeong, ``Nonclassical light and metrological power: An
  introductory review,'' \emph{AVS Quantum Science}, vol.~1, no.~1, p. 014701,
  2019.

\bibitem{herrero2017quantum}
M.~Herrero-Collantes and J.~C. Garcia-Escartin, ``Quantum random number
  generators,'' \emph{Rev. Mod. Phys.}, vol.~89, no.~1, p. 015004, 2017.

\bibitem{braun2018quantum}
D.~Braun, G.~Adesso, F.~Benatti, R.~Floreanini, U.~Marzolino, M.~W. Mitchell,
  and S.~Pirandola, ``Quantum-enhanced measurements without entanglement,''
  \emph{Rev. Mod. Phys.}, vol.~90, no.~3, p. 035006, 2018.

\bibitem{barsotti2018squeezed}
L.~Barsotti, J.~Harms, and R.~Schnabel, ``Squeezed vacuum states of light for
  gravitational wave detectors,'' \emph{Rep. Progr. Phys.}, vol.~82, no.~1, p.
  016905, 2018.

\bibitem{o2009photonic}
J.~L. O'Brien, A.~Furusawa, and J.~Vu{\v{c}}kovi{\'c}, ``Photonic quantum
  technologies,'' \emph{Nat. Phot.}, vol.~3, no.~12, pp. 687--695, 2009.

\bibitem{horodecki2009quantum}
R.~Horodecki, P.~Horodecki, M.~Horodecki, and K.~Horodecki, ``Quantum
  entanglement,'' \emph{Rev. Mod. Phys.}, vol.~81, no.~2, p. 865, 2009.

\bibitem{korolkova2019quantum}
N.~Korolkova and G.~Leuchs, ``Quantum correlations in separable multi-mode
  states and in classically entangled light,'' \emph{Rep. Prog. Phys.},
  vol.~82, no.~5, p. 056001, 2019.

\bibitem{yu2009sudden}
T.~Yu and J.~Eberly, ``Sudden death of entanglement,'' \emph{Science}, vol.
  323, no. 5914, pp. 598--601, 2009.

\bibitem{weedbrook2016discord}
C.~Weedbrook, S.~Pirandola, J.~Thompson, V.~Vedral, and M.~Gu, ``How discord
  underlies the noise resilience of quantum illumination,'' \emph{New J.
  Phys.}, vol.~18, no.~4, p. 043027, 2016.

\bibitem{kay1993fundamentals}
S.~M. Kay, \emph{Fundamentals of statistical signal processing}.\hskip 1em plus
  0.5em minus 0.4em\relax Prentice Hall PTR, 1993.

\bibitem{petz2011introduction}
D.~Petz and C.~Ghinea, ``Introduction to quantum {F}isher information,'' in
  \emph{Quantum probability and related topics}.\hskip 1em plus 0.5em minus
  0.4em\relax World Scientific, 2011, pp. 261--281.


\bibitem{lanzanew} M. J. Brandsema, M. Lanzagorta and R. M. Narayanan, "Equivalence of Classical and Quantum Electromagnetic Scattering in the Far-Field Regime," IEEE Aerospace and Electronic Systems Magazine, \textbf{35}, pp. 58-73, 2020. 


\bibitem{mikhalychev2018synthesis}
A.~Mikhalychev, D.~Mogilevtsev, G.~Y. Slepyan, I.~Karuseichyk, G.~Buchs,
  D.~Boiko, and A.~Boag, ``Synthesis of quantum antennas for shaping field
  correlations,'' \emph{Phys. Rev. App.}, vol.~9, no.~2, p. 024021, 2018.

\bibitem{shapiro2009quantum}
J.~H. Shapiro and S.~Lloyd, ``Quantum illumination versus coherent-state target
  detection,'' \emph{New J. Phys.}, vol.~11, no.~6, p. 063045, 2009.

\bibitem{tan2008quantum}
S.-H. Tan, B.~I. Erkmen, V.~Giovannetti, S.~Guha, S.~Lloyd, L.~Maccone,
  S.~Pirandola, and J.~H. Shapiro, ``Quantum illumination with gaussian
  states,'' \emph{Phys. Rev. Lett.}, vol. 101, no.~25, p. 253601, 2008.

\bibitem{guha2009gaussian}
S.~Guha and B.~I. Erkmen, ``Gaussian-state quantum-illumination receivers for
  target detection,'' \emph{Phys. Rev. A}, vol.~80, no.~5, p. 052310, 2009.

\bibitem{zhuang2017optimum}
Q.~Zhuang, Z.~Zhang, and J.~H. Shapiro, ``Optimum mixed-state discrimination
  for noisy entanglement-enhanced sensing,'' \emph{Phys. Rev. Lett.}, vol. 118,
  no.~4, p. 040801, 2017.



\bibitem{messaoudi2020quantum}
N.~Messaoudi, C.~W. Chang, A.~Vadiraj, J.~Bourassa, B.~Balaji, and C.~Wilson,
  ``Quantum-enhanced noise radar,'' \emph{Appl. Phys. Lett.},  vol.~ 114, p. 112601, 2019.

\bibitem{zhuang2017quantum}
Q.~Zhuang, Z.~Zhang, and J.~H. Shapiro, ``Quantum illumination for enhanced
  detection of {R}ayleigh-fading targets,'' \emph{Phys. Rev. A}, vol.~96,
  no.~2, p. 020302, 2017.

\bibitem{peshko2019quantum}
I.~Peshko, D.~Mogilevtsev, I.~Karuseichyk, A.~Mikhalychev, A.~Nizovtsev, G.~Y.
  Slepyan, and A.~Boag, ``Quantum noise radar: superresolution with quantum
  antennas by accessing spatiotemporal correlations,'' \emph{Opt. Exp.},
  vol.~27, no.~20, pp. 29\,217--29\,231, 2019.

\bibitem{rayleigh1879xxxi}
L.~Rayleigh, ``Investigations in optics, with special reference to the
  spectroscope,'' \emph{The London, Edinburgh, and Dublin Philosophical
  Magazine and Journal of Science}, vol.~8, no.~49, pp. 261--274, 1879.

\bibitem{tsang2016quantum}
M.~Tsang, R.~Nair, and X.-M. Lu, ``Quantum theory of superresolution for two
  incoherent optical point sources,'' \emph{Phys. Rev. X}, vol.~6, no.~3, p.
  031033, 2016.


\bibitem{li2020super}
Z.-P. Li, X.~Huang, P.-Y. Jiang, Y.~Hong, C.~Yu, Y.~Cao, J.~Zhang, F.~Xu, and
  J.-W. Pan, ``Super-resolution single-photon imaging at 8.2 kilometers,''
  \emph{Opt. Exp.}, vol.~28, no.~3, pp. 4076--4087, 2020.

\bibitem{drummond2013}
M. Hillery and  P.~D. Drummond, \emph{The Quantum Theory of Nonlinear Optics}.\hskip 1em
  plus 0.5em minus 0.4em\relax Cambridge University Press, 2014.

\bibitem{solntsev2017path}
A.~S. Solntsev and A.~A. Sukhorukov, ``Path-entangled photon sources on
  nonlinear chips,'' \emph{Rev. Phys.}, vol.~2, pp. 19--31, 2017.

\bibitem{makhlin2001quantum}
Y.~Makhlin, G.~Sch{\"o}n, and A.~Shnirman, ``Quantum-state engineering with
  {J}osephson-junction devices,'' \emph{Rev. Mod. Phys.}, vol.~73, no.~2, p.
  357, 2001.

\bibitem{zhong2013squeezing}
L.~Zhong, E.~Menzel, R.~Di~Candia, P.~Eder, M.~Ihmig, A.~Baust, M.~Haeberlein,
  E.~Hoffmann, K.~Inomata, T.~Yamamoto \emph{et~al.}, ``Squeezing with a
  flux-driven {J}osephson parametric amplifier,'' \emph{New J. Phys.}, vol.~15,
  no.~12, p. 125013, 2013.

\bibitem{barzanjeh2020microwave}
S.~Barzanjeh, S.~Pirandola, D.~Vitali, and J.~Fink, ``Microwave quantum
  illumination using a digital receiver,'' \emph{Sci. Adv.}, vol.~6, no.~19, p.
  eabb0451, 2020.

\bibitem{Marino:19}
G.~Marino, A.~S. Solntsev, L.~Xu, V.~F. Gili, L.~Carletti, A.~N. Poddubny,
  M.~Rahmani, D.~A. Smirnova, H.~Chen, A.~Lema\^{i}tre, G.~Zhang, A.~V. Zayats,
  C.~D. Angelis, G.~Leo, A.~A. Sukhorukov, and D.~N. Neshev, ``Spontaneous
  photon-pair generation from a dielectric nanoantenna,'' \emph{Optica},
  vol.~6, no.~11, pp. 1416--1422, 2019.

\bibitem{luong2020quantum}
D.~Luong and B.~Balaji, ``Quantum two-mode squeezing radar and noise radar,''
  \emph{IET Radar, Sonar \& Navigation}, vol.~14, no.~1, pp. 97--104, 2020.

\bibitem{french} G. Sorelli, N. Treps, F. Grosshans, F. Boust, Detecting a target with quantum entanglement. 2020. ⟨hal-02877841⟩.

\end{thebibliography}
\end{document}